# mHealth Strategy to Fight Tuberculosis in Bangladesh

Md Monzur Morshed
Department of Development Studies
University of Dhaka[1]
TigerHATS[2]
m.monzur@gmail.com, monzur@tigerhats.org

## Executive Summary

Bangladesh is one of the high TB burden countries in the world and TB is still a major public health problem in the country. To eradicate TB in Bangladesh and to ensure proper monitoring and better health care service, digital interventions may play a pivotal role. Over the years Bangladesh Government has been actively working to hold a control over TB with the help of donors, local and international NGOs. For sustainability there has been a need to develop mHealth strategy for Bangladesh to align with the Digital Bangladesh Vision 2021.

## Bangladesh Context (gaps and strengths)

In Digital Bangladesh Vision 2021, the Government of Bangladesh put emphasis on ICT-based intervention in health sector and placed a high priority on e-Health to build the capacity and management strength of the healthcare delivery system to ensure citizens access to quality health care service and to be prepared for emerging health threats (such as TB, HIV, COVID19) and challenges [1].

| Strengths | Weaknesses |
|---|---|
| <ul><li>Willingness to Digitize</li><li>Budget</li><li>Technical expertise</li><li>Scope to implement</li></ul> | <ul><li>Lack of human resource</li><li>Fearfulness to adopt modern technology</li><li>Lack of monitoring</li><li>Bureaucratic</li><li>Lack of advocacy</li></ul> |
| **Opportunities** | **Threats** |
| <ul><li>Available data</li><li>Available skilled human resource</li><li>Willingness of the donors</li></ul> | <ul><li>Funding dependency</li><li>Lack of sustainability</li><li>Replication of similar ideas</li></ul> |

SWOT Analysis of Bangladesh Health Sector in terms of Digitization

In 4th Health, Population & Nutrition Sector Programme (4th HPNSP) operational plan for the implementation period of January 2017-June 2022, government of Bangladesh has identified the objective to improve health information system, eHealth and medical biotechnology. To be more specific, there is ample opportunity to contribute in eHealth by continuation and further development of health call center, mobile phone health service and other mHealth.

**Intervention**

Over the years USAID has been supporting Bangladesh in health sector and supporting mHealth based innovations. Digital Solutions such as mHealth infrastructure may work as a catalyst to fight infectious diseases like TB, COVID19. According to the WHO's TB Digital Strategy, mHealth solutions are categorized into four areas such as i) patient care, ii) surveillance and monitoring, iii) programme management and iv) eLearning. Since Bangladesh Government is very sincere to digitize the entire health sector and to reach grass root, a comprehensive mHealth strategy would be a game changer.

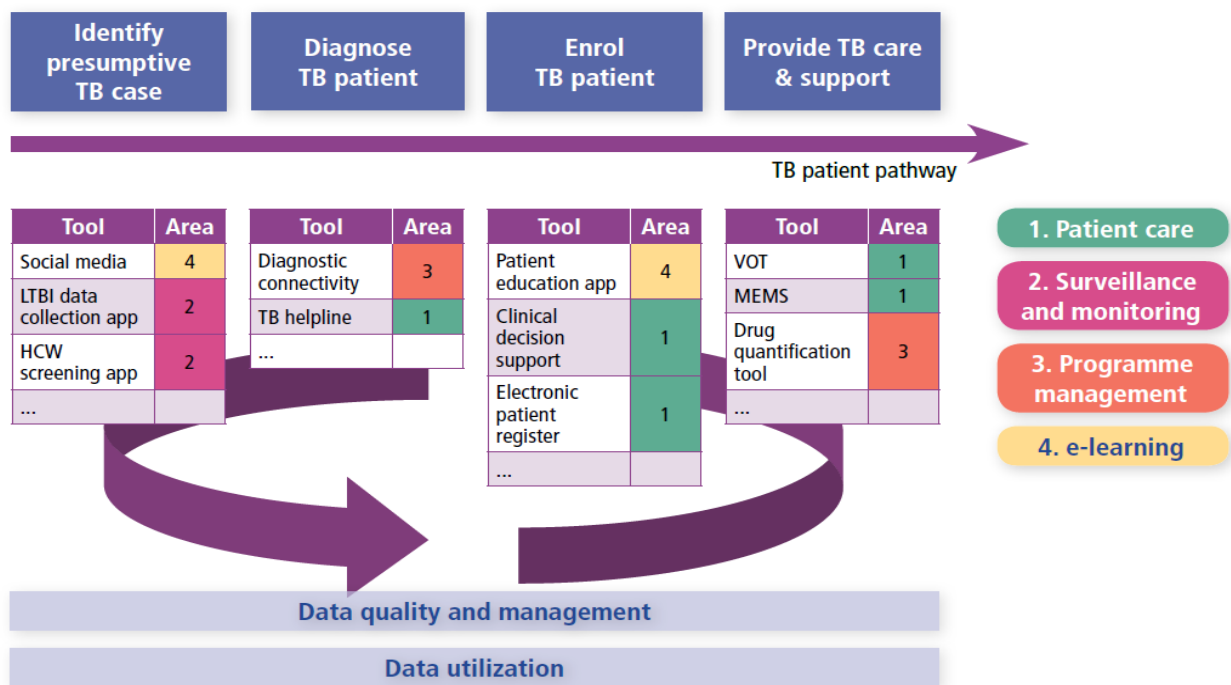

HCW: health-care worker; LIMS: laboratory information management systems; LTBI: latent tuberculosis infection; MEMS: medication event monitoring system; VOT: video-supported treatment for TB

Following: Handbook for the Use of Digital Technologies to Support TB Medication Adherence. WHO 2017. http://apps.who.int/iris/bitstream/handle/10665/259832/9789241513456-eng.pdf

# Planning for Implementation

Concept development plays a crucial role in mHealth solution development. In this phase every stakeholder shall be involved and definitely it shall be bottom up approach. Understanding the local context is important for any mHealth solution. Field research such as surveying and interviewing local people, health workers, Government health officials, doctors, decision makers, etc. would help to identify the requirements precisely. Several workshops and joint need assessment sessions can be organized to gather the requirements in a scientific way.

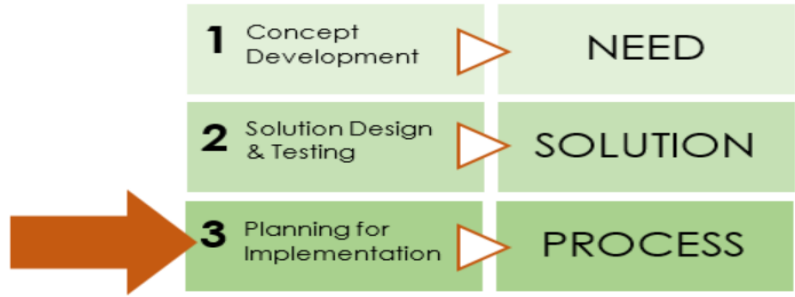

Following: Planning for Implementation, https://www.k4health.org/toolkits/mhealth-planning-guide/planning-implementation

Based on the research findings, a standard documentation shall be drafted to baseline the requirements that will highlight the blue print of the functional and non-functional features of the mHealth applications. Before going into the final implementation phase, it is recommended to approach for piloting of the mHealth applications. Often it has been seen that stakeholders do not put priority on the piloting, instead they focus on the direction implementation which may jeopardize the entire mission.

| Strength | Weakness |
|---|---|
| <ul><li>Widely accepted</li><li>Open source</li><li>Cross Platform</li><li>Scalable</li><li>Community support</li><li>Documentation</li></ul> | <ul><li>Unstable platform</li><li>Unstable technology</li><li>Too many bugs/defects</li><li>Too much dependency</li><li>Lack of documentation</li></ul> |
| Opportunity | Threat |
| <ul><li>Competitive price</li><li>Available skilled human resource</li><li>Popular technology</li></ul> | <ul><li>Structural change leads to affect the core technology</li><li>Third party software dependency</li><li>Copyright issue</li><li>Unrealistic price</li><li>Security</li></ul> |

Following: SWOT Analysis in mHealth Architecture Decision-Making [5] (M. Morshed, 2019), https://link.springer.com/chapter/10.1007/978-981-13-6861-5_1

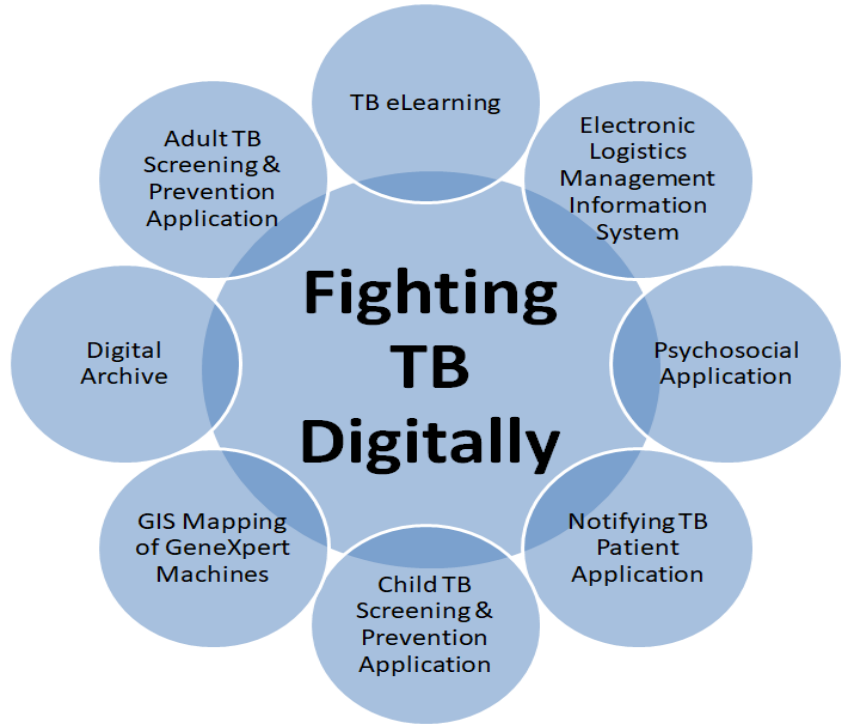

Following: mHealth Solutions adopted by Challenge TB (2015-2019), USAID Funded Project

## Recommendations

- A joint community/ consortium can be formed with the participation of Government, NGO, and Donors which may support in further advocacy and communication.
- A common social network platform can be established to strengthen the communication with the related stakeholders.
- Providing sensitization on the mHealth Solutions may influence the decision makers to realize the need.
- Joint workshop can be organized and to be participated by influential personals, policy makers and government officials from DGHS, NTP, MoHFW, etc.
- Utilizing Social media, TV to inform people about TB and promoting mHealth solutions.
- Adoption of telemedicine (such as IVR platform) could bring a new dimension.
- Need to take new initiatives to strengthen the e-Health infrastructure and building the human capacity of the health workers.

# Conclusion

Sustainability of the mHealth Solutions is one of the major concerns for stakeholders like NTP, DGHS, DGFP, implementation partners and donors. New innovations are happening every year, but not in a collaborative way. Most often it has been seen that similar types of innovations have been replicated. To overcome these challenges an open knowledge and communication forum can be established to strengthen the health communication and capacity of the Bangladesh Health Sector.

# References

[1] https://a2i.gov.bd/wp-content/uploads/2017/11/4-Strategy_Digital_Bangladesh_2011.pdf
[2] http://www.dghs.gov.bd/images/docs/OP/2018/E-Health%20(HIS-e-H).pdf
[3] http://apps.who.int/iris/bitstream/handle/10665/259832/9789241513456-eng.pdf
[4] https://www.k4health.org/toolkits/mhealth-planning-guide/planning-implementation
[5] Morshed M.M., Hasan M., Rokonuzzaman M. (2019) Software Architecture Decision-Making Practices and Recommendations. In: Bhatia S., Tiwari S., Mishra K., Trivedi M. (eds) Advances in Computer Communication and Computational Sciences. Advances in Intelligent Systems and Computing, vol 924. Springer, Singapore